\documentclass{article}

\usepackage{arxiv}

\usepackage[utf8]{inputenc} 
\usepackage[T1]{fontenc}    
\usepackage{hyperref}       
\usepackage{url}            
\usepackage{booktabs}       
\usepackage{amsfonts}       
\usepackage{nicefrac}       
\usepackage{microtype}      
\usepackage{textcomp}
\usepackage{graphicx}
\usepackage{lipsum}

\title{Virtual replicas of real places: Experimental investigations}

\author{
  Richard Skarbez\thanks{Dr. Skarbez is now at La Trobe University and can be reached at r.skarbez@latrobe.edu.au.} \\
  Center for Human-Computer Interaction\\
  Virginia Tech\\
  Blacksburg VA \\
  \texttt{rskarbez@vt.edu} \\
   \And
  Doug A. Bowman \\
  Center for Human-Computer Interaction\\
  Virginia Tech\\
  Blacksburg VA \\
  \texttt{dbowman@vt.edu} \\
  \And
  J. Todd Ogle \\
  Center for Human-Computer Interaction\\
  Virginia Tech\\
  Blacksburg VA \\
  \texttt{jogle@vt.edu} \\
  \And
  Thomas Tucker \\
  Center for Human-Computer Interaction\\
  Virginia Tech\\
  Blacksburg VA \\
  \texttt{thomasjt@vt.edu} \\  
  \And
  Joseph L. Gabbard \\
  Center for Human-Computer Interaction\\
  Virginia Tech\\
  Blacksburg VA \\
  \texttt{jgabbard@vt.edu} \\  
}

\begin{document}
\maketitle

\begin{abstract}
The emergence of social virtual reality (VR) experiences, such as Facebook Spaces, Oculus Rooms, and Oculus Venues, will generate increased interest from users who want to share real places (both personal and public) with their fellow users in VR. At the same time, advances in scanning and reconstruction technology are making the realistic capture of real places more and more feasible. These complementary pressures mean that the representation of real places in virtual reality will be an increasingly common use case for VR. Despite this, there has been very little research into how users perceive such replicated spaces. This paper reports the results from a series of three user studies investigating this topic. Taken together, these results show that getting the scale of the space correct is the most important factor for generating a ``feeling of reality'', that it is important to avoid incoherent behaviors (such as floating objects), and that lighting makes little difference to perceptual similarity.
\end{abstract}


\section{Introduction}
In recent years, there have been a number of articles claiming that social virtual reality (VR) is, or will be, the ``killer app'' for virtual reality. (Some titles include: ``How Facebook's social VR could be the killer app for virtual reality,'' ``Virtual reality's killer app will be talking to your friends,'' ``Facebook Spaces is the killer app for VR and its first conquest might be Twitch,'' ``Why co-experience is the ultimate killer app for virtual reality,'' and ``Will the killer app for VR be shared live experiences?'' \cite{Terdiman:2016:FacebookSocialVR} \cite{Smith:2015:KillerApp} \cite{Ganz:2017:FacebookSpaces} \cite{Baszucki:2017:CoExperience} \cite{Bonasio:2017:SharedLiveExperiences}) VR applications such as Facebook Spaces\footnote{https://www.facebook.com/spaces}, Oculus Rooms\footnote{https://www.oculus.com/experiences/go/1101959559889232/}, and Oculus Venues\footnote{https://www.oculus.com/experiences/go/1555304044520126/} enable users to spend time in shared virtual spaces. At the moment these spaces are most often purely virtual, but offer personal customization. It seems likely that rather than sharing a virtual space with limited personalization, users might prefer to inhabit a virtual replica of a meaningful real space, such as their own living room. At the moment, 3D scanning has not been widely adopted by home users, but products such as the Microsoft Kinect\footnote{https://en.wikipedia.org/wiki/Kinect} and Cubify Sense\footnote{https://www.3dsystems.com/shop/sense} have demonstrated that 3D scanning at consumer-level prices is increasingly feasible.

Widespread adoption of these technologies would have applications beyond social VR as well. Other application domains that would benefit from the ability to replicate real places in VR include cultural heritage, real estate, and architecture.

For these reasons, we believe that creating and inhabiting replicas of real places will be an increasingly common use case for VR. However, to date, there has been very little research into how users perceive and interact with such replicated spaces. Some important questions in this area include: How accurate do replica spaces need to be? Are there some elements of replica spaces for which accuracy is more important than others? Are there elements of a VR space that generally go unnoticed, and thus do not need significant technical investment in terms of scanning and reconstruction? With this paper, we hope to contribute some useful knowledge in this area. We report on the design and results of a series of three user studies which had the goal of determining which characteristics of virtual rooms were most important for users to have the same ``feeling of reality'' as in the original real room. (The ultimate goal of the project was to produce a perceptual similarity measure that could be used to evaluate the quality of virtual replicas of a given real space. This paper focuses of the design and results of the studies that were intended to inform the creation of that measure. The design and implementation of the measure itself are beyond the scope of this paper.) Results from these studies include that the scale of the room and large objects in it are most important for users to perceive the room as real, that non-physical behaviors such as objects floating in air are readily noticeable and have a negative effect even when the errors are small in scale, and that differences in lighting quality seem to have a minimal effect on users' perception of a replica space.

\section{Previous Work}
\label{sec:previousWork}

``Feeling of reality'' can be mapped onto existing constructs, such as Ba\~{n}os \emph{et al.}'s Reality Judgment \cite{Banos:2000:RJPQ} or Slater's Plausibility Illusion (Psi) \cite{Slater:2009:PIandPsi}. We adopt Slater's term, which he defines as, ``the illusion that what is apparently happening is really happening (even though you know for sure that it is not).'' Skarbez, Brooks, and Whitton argue that Plausibility Illusion derives from a construct they call coherence, which is essentially the extent to which a scenario complies with a user's expectations \cite{Skarbez:2017:PresenceSurvey}. (Elsewhere, Gilbert uses the term authenticity for essentially the same construct \cite{Gilbert:2017:PerceivedRealism}.) Per Skarbez, ``Coherence can be thought of as a superset of realism or fidelity. Specifically, coherence makes no assumptions about a VE having to faithfully represent the real world. Rather, coherence depends on the internal logical and behavioral consistency of the virtual experience.''

Since the scenario we are presenting to users is that they are inhabiting a virtual replica of a real room, prior knowledge is all of users' previous experiences with the real world. Therefore, Plausibility Illusion (or ``feeling of reality'') should be negatively affected by any characteristics of the virtual environment that are not consistent with experience in the real world. Some examples of this could include: objects floating above the ground (inconsistent with prior experience with gravity), objects interpenetrating one another, models containing holes, models being represented by ``billboards'', objects appearing ``tessellated'', models having noticeably low texture resolution (inconsistent with prior experience with physical objects), walls not being flat/corners not being square, traditionally planar surfaces (such as walls) appearing to have bends or dips, room dimensions being too big or too small, doors not being ``door-shaped'' (inconsistent with prior experience of the built world), objects being significantly moved, or not looking like the original objects (inconsistent with prior experience with the real original room).

There has been little research into the effect of ``realism'' on user experience in virtual environments, and what has been done to date has primarily focused on its impact on presence. Bouchard \emph{et al.} demonstrated that participants' belief that the scenario represented the real world, as opposed to being a virtual recreation, resulted in higher presence scores \cite{Bouchard:2012:SubjectiveRealism}. Hvass \emph{et al.} describe a significant (but small) effect of geometric realism (polygon count + texture resolution) on physiological and questionnaire-based measures of presence \cite{Hvass:2017:GeometricRealism}. Slater \emph{et al.} argue that increased realism (real-time ray tracing as opposed to ray casting) increased stress in a stressful virtual environment, and therefore increased presence \cite{Slater:2009:VisualRealismPsi}. However, a follow-up study indicates that the increase in presence was due to the addition of dynamic behavior to the environment (shadows and reflections), rather than the illumination quality itself \cite{Yu:2012:VisualRealism}. Welch \emph{et al.} found that changing pictorial realism had a significant effect on presence, but anecdotally, the effect was less important than that of interactivity or latency \cite{Welch:1996:PictorialRealism}. (The experiment was a driving task; pictorial realism was manipulated by changing the color of the sky from blue to black, the landscape from hilly to flat, the background from green to black, and removing peripheral objects and oncoming cars. It can be argued, and the authors acknowledge, that most of these changes may confound scene complexity with scene realism.) 	

That said, there have been some investigations into the effects of visual realism on other aspects of user experience and behavior. Thompson \emph{et al.} investigated the effects of visual realism on distance (under)estimation in virtual environments, and found that increasing the realism of the virtual scene had no significant effect on distance underestimation \cite{Thompson:2004:ComputerGraphics}. Lee \emph{et al.} explored the effects of visual realism on performance of search tasks in mixed reality environments, and found no significant effect of visual realism on task performance or on presence \cite{Lee:2013:VisualRealism}. Ragan \emph{et al.} explored the effects of visual complexity on performance and training transfer in a visual search task, and found that high visual complexity led to worse performance during the training trials, but better adherence to the search strategy during the evaluation trials \cite{Ragan:2015:VisualScanningTask}.

\section{Overview of user studies}
\label{sec:experimentalDesign}

 
The first user study, described in Section \ref{sec:experiment1}, was a pilot study of which the goals were to: demonstrate the feasibility of our virtual experimental testbed, check whether any parameters important to users were not considered in the creation of the testbed, generate a ``first pass'' ranking of the importance of various parameters, and most critically, identify which parameters merited further investigation in User Studies 2 and 3.
 
The second study, described in Section \ref{sec:experiment2}, was a classical psychophysical study designed to evaluate the subset of parameters identified in Study 1 with respect to one another. Specifically, this user study sought to identify perceptual equivalences between different parameters so that the parameters could be correctly valued in the budget-based Study 3.
 
The third user study, described in Section \ref{sec:experiment3}, used a budget-based method derived from the method introduced by Slater in \cite{Slater:2010:SimulatingVEsInVEs} and employed by Azevedo, Jorge, and Campos \cite{Azevedo:2014:EEGData}, Bergstr\"{o}m \emph{et al.} \cite{Bergstrom:2017:StringQuartet}, and Skarbez \emph{et al.} \cite{Skarbez:2017:PsychophysicalExperiment}. In this study, participants were presented with the replica environment in some initial (degraded) state, and were given the opportunity to upgrade the environment through manipulation of parameters (the same from Study 2) in whichever order they saw fit and to whichever extent they saw fit, given an overall budget constraint. This user study sought to establish which parameters participants considered most important, and ``how correct'' those parameters needed to be to satisfy participants.

\section{User study 1 (Pilot)}
\label{sec:experiment1}

As described in Section \ref{sec:experimentalDesign}, this was a pilot study of which the primary goal was to identify parameters that required further investigation. Thus, we sought to investigate a large set of parameters that we could confidently winnow down for use in subsequent user studies.

\begin{figure}[htbp]
 \centering 
 \includegraphics[width=\textwidth]{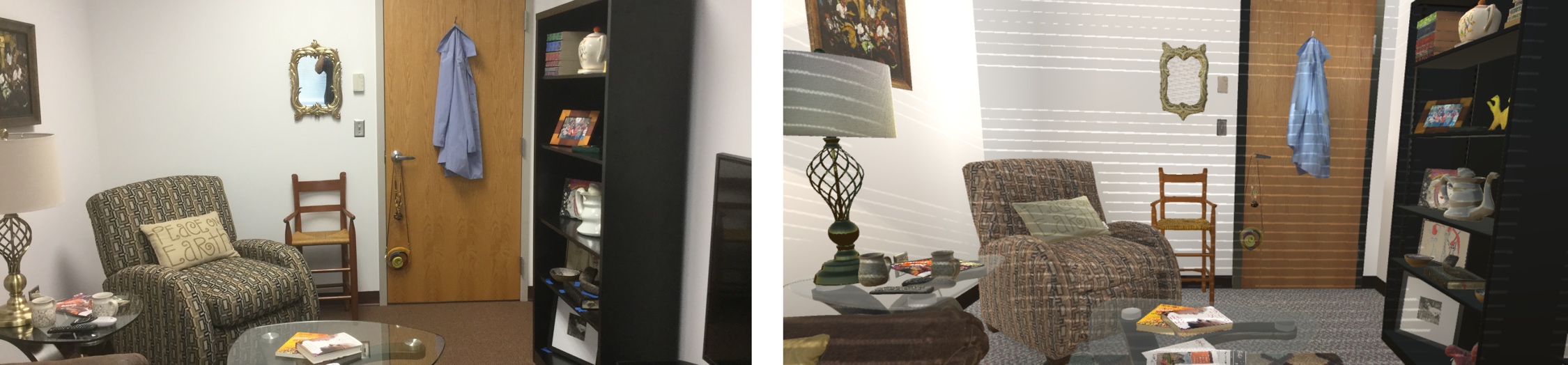}
 \caption{Real original room at left, virtual replica room at right.}
 \label{fig:teaser}
\end{figure}

In order to investigate which parameter changes were perceptually important to participants, modified replica rooms were created by changing several parameters of the ``ground truth'' (GT) room---the most accurate version of the virtual replica room, as seen in Figure \ref{fig:teaser} (right). After significant iterative internal testing with the team, we decided to alter most parameters by $\pm$10\%, $\pm$25\%, $\pm$50\%, or $\pm$75\% depending upon the parameter. The goal was to create obvious, noticeable differences between parameter levels for this study, so as to evoke strong responses from Study 1 participants, knowing that in future studies participants would be able to manipulate the parameters with fine granularity. Modifications were drawn from the following list:
\begin{itemize}
	\item \textbf{Ceiling height} changed by +25\%, +10\%, -5\%, and -10\%. (The asymmetry is due to the fact that internal testing revealed that for some users, -25\% put their head at ceiling level, which was immediately noticeable and disturbing.)
	\item \textbf{Room length} changed by +25\%, +10\%, -10\%, and -25\%.
	\item \textbf{Room width} changed by +25\%, +10\%, -10\%, and -25\%.
	\item \textbf{Door removed}. The door was replaced by a blank wall.
	\item \textbf{Window removed}. The window was replaced by a blank wall.
	\item \textbf{Wall material} changed in color or texture.
	\item \textbf{Furniture removed}. One of the following was removed at a time: couch, coffee table, side tables, TV stand, and a stuffed chair. When a piece of furniture was removed, all clutter that was ``on'' that furniture object was also removed, to eliminate the incoherent stimulus of floating objects.
	\item \textbf{Furniture quality reduced}. For both the couch and the stuffed chair, models were generated that contained 10\%, 25\%, 50\%, and 75\% of the original number of vertices. If ``decimated furniture'' was one of the errors in a given room, the sofa and the stuffed chair always varied together.
	\item \textbf{Furniture mismatched}. One of the following was replaced at a time: coffee table, side tables, stuffed chair, couch, lamps, and a wooden chair. When a piece of furniture was mismatched, it was replaced with another similar object of the same type taken from the ShapeNet database \cite{Chang:2015:ShapeNet}.
	\item \textbf{Furniture repositioned}. Furniture objects were either globally raised by 10cm, globally lowered by 10cm, or globally moved outward (away from room center) by 10\%. The corresponding ``moved inward by 10\% condition'' was not tested due to experimenter error.
	\item \textbf{Furniture rescaled}. One of the following scaling errors occurred: All furniture was 25\% larger, all furniture was 10\% larger, all furniture was 10\% smaller, the sofa was 25\% larger, the sofa was 10\% smaller, the coffee table was 25\% larger. When a furniture object was scaled up, all clutter that was ``on'' that furniture object was also scaled up, to avoid giving participants conflicting context cues.
	\item \textbf{Clutter removed}. Either 10\% (9) of clutter objects were hidden, 25\% (22) of clutter objects were hidden, or 50\% (44) clutter objects were hidden. Clutter objects were hidden at random, and in different random orders for each of the three conditions. Physical constraints were not enforced; for example, it was possible for the middle book in a stack of three to be hidden.
	\item \textbf{Clutter quality reduced}. For the mask, the white pitcher, the bongos, and the pig statue, models were generated that contained 10\%, 25\%, 50\%, and 75\% of the original number of vertices. If ``decimated clutter'' was one of the errors in a given room, all four objects always varied together.
	\item \textbf{Clutter repositioned}. One of the following positioning errors occurred: Four objects---always the dumbbells, the coffee mug, the ISE brochure, and the white pitcher---were elevated by 10cm, the four objects were elevated by 5cm, the four objects were lowered by 5cm, the four objects were lowered by 10cm, all clutter was moved inward by 25\%, or all clutter was moved outward by 10\%.
	\item \textbf{Clutter rescaled}. One of the following scaling errors occurred: the four objects were  25\% larger, the four objects were 10\% larger, the four objects were 10\% smaller, the four objects were 25\% smaller, all clutter was 25\% larger, all clutter was 10\% larger, all clutter was 10\% smaller, all clutter was 25\% smaller.
	\item \textbf{Lights missing}. One of the following lights was ``turned off'': ``sunlight'', ceiling lights, or both lamps. Note that ``sunlight'' was implemented as a directional light outside the room, and the ceiling lights did not actually light the room, since area lights could not be enabled---turning the ceiling lights on or off only changed whether the ceiling light panels appeared to ``glow'' or not.
	\item \textbf{Light brightness changed}. All lights were either 4 times as bright, 2 times as bright, \textonehalf\ as bright, or \textonequarter\ as bright. All lights always varied together.
	\item \textbf{Light color changed}. All lights were interpolated 25\% toward blue, 10\% toward blue, 10\% toward red, or 25\% toward red. All lights, including the ambient lighting, always varied together.
\end{itemize}
 
\begin{figure}[htbp]
 \centering 
 \includegraphics[width=\textwidth]{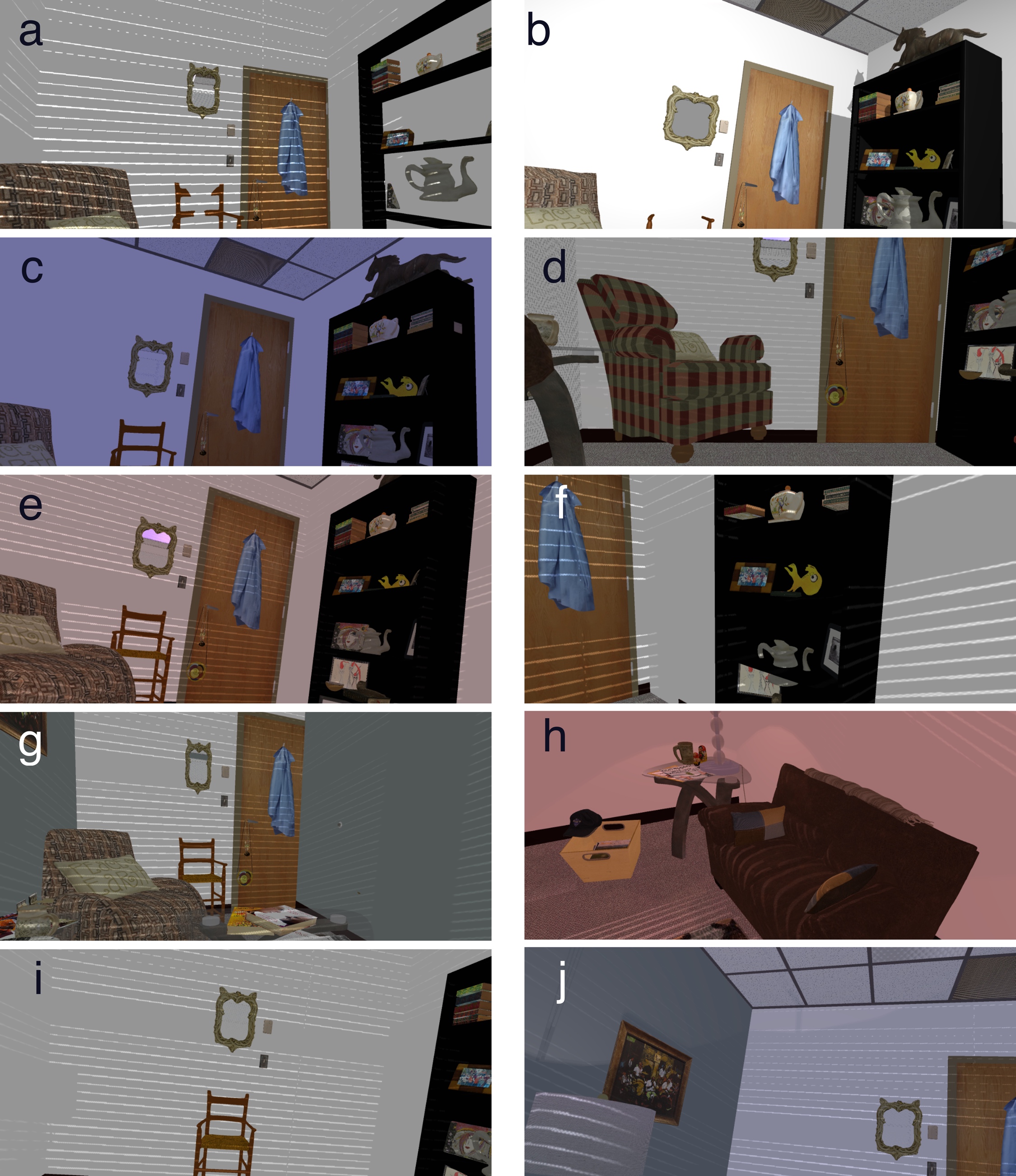}
 \caption{Screenshots from each of the ten modified replica rooms participants experienced in Study 1.}
 \label{fig:study1_combined}
\end{figure}

In total, ten different modified replica rooms were created, each of which had between five and nine parameters modified. Screenshots for each of these rooms appear in Figure \ref{fig:study1_combined}.

\subsection{Participants}
\label{sec:experiment1_participants}

Six participants were recruited from the students, faculty, and staff of Virginia Tech's Grado Department of Industrial and Systems Engineering, in which the study was being conducted. For this study, participant descriptors such as age, gender, etc. were not recorded, and participants were not compensated.
 
\subsection{Materials}
\label{sec:experiment1_materials}

For this research, we staged a full-size actual room with furnishings typical of a living room including a sofa, two chairs, end tables, coffee table, lamps, a television, bookshelf, art, and several other smaller items. We chose to model a living room since this is a common place not only in most homes, but also a space whereby remote, VR-enabled social interactions could be likely. We added clutter to many of the horizontal surfaces (e.g., coffee and end tables, bookshelves, the floor) that would representative in a real living room setting.  These clutter included a set of bongos, a metal pig sculpture, free standing picture frames, books, pamphlets, remote controls, drinking cups, small toys, and many others. In all there were just over 100 items in the room.

The virtual environment was displayed using an Oculus Rift CV1 head-worn display (HWD). The Rift has a nominal 110\textdegree\, field of view, and a resolution of 1080x1200 pixels per eye. It weighs 470 grams.
 
For tracking, the Rift tracking system was used in a roomscale (3 camera) configuration. The size of the tracked space was approximately 2m x 3m, and was comparable to the navigable space available in the real room.
 
The experimental testbed was implemented using version 5.6 of the Unity game engine. The base room model was generated using a combination of laser scanning to obtain global information, photogrammetry scanning of individual small objects, and additional processing using Autodesk Maya and 3ds Max. A comparison of a photograph of the real room and a screenshot of the virtual replica environment can be seen in Figure \ref{fig:teaser}.
 
\subsection{Measures}
\label{sec:experiment1_measures}

For this study, participants were asked to ``think aloud'' as they explored the environment, and specifically to comment on things that seemed unusual about the virtual environment. These comments were transcribed by an experimenter, and these comments were analyzed in order to generate descriptive statistics regarding (1) what the ``unusual'' things that were noticed by participants were, and (2) in what order they were commented upon.

\subsection{Procedures}
\label{sec:experiment1_procedures}

Upon arriving at the lab, participants received a brief description of the task, and then donned the Rift HWD. Participants first experienced the GT room; this was done so that participants would have a baseline experience against which to compare the subsequent degraded virtual rooms. Participants were then told that they would experience a series of modified versions of these rooms, and that they should ``think aloud'' and comment on the differences they noticed, if any, between the modified room and the GT room. Participants were re-exposed to the GT room after each exposure to a modified room so as to refresh their memory. The entire session lasted approximately 30 minutes.

\subsection{Results}
\label{sec:experiment1_results}

In this section, we summarize the results of this user study by listing the parameters that were not selected for further investigation in Studies 2 and 3, followed by those that were, with associated justifications.

\subsubsection{Parameters not selected for further investigation}
\label{sec:experiment1_results_not_included}

 
\textbf{Ceiling Height} errors were not noticed by any participants at +10\%, -5\%, and -10\%, and were only noticed by some participants at +25\%. This can potentially be explained in several ways. First, the vast majority of objects in the room (and in fact in most real-world rooms) were at eye-level or below, so there was no compelling need to look up. Second, it has previously been observed that wearing an HWD decreases the amount of vertical head motion \cite{Keller:1998:TerribleHMDs}, probably due to the fact that it applies a weight to the front of the head, making raising the head more difficult. For these reasons, we expected that ``reasonable'' errors in ceiling height would not be noticed by participants. 
 
\textbf{Missing Door} and \textbf{Missing Window} errors were always noticed by all participants, with early ``noticing orders.'' Therefore, we saw no need to further investigate these errors. 
 
\textbf{Missing Furniture} errors were almost always noticed by all participants, with early noticing orders. As with missing doors/windows, then, missing furniture errors were not investigated further.
 
\textbf{Furniture Translation} errors were mostly not noticed by participants. It makes sense that people are less sensitive to translation errors than elevation errors, as even gross translation errors are still physically plausible. 
 
\textbf{Clutter Translation} errors were mostly not noticed by participants. The same arguments regarding Furniture Translation also apply here. 
 
\textbf{Clutter Rescaling} errors were generally noticed by some participants, but never by all participants. Participants were much less sensitive to these errors than to Furniture Rescaling errors. This makes sense on face: furniture objects take up more space on the retina, so drawing more attention; additionally, a 1\% difference in scale corresponds to a greater absolute change for large objects than for small ones. 
 
 
\textbf{Furniture Quality} errors were never noticed at 25\% and 75\% of vertices (and probably not at 50\% of vertices; we suspect this is a spurious result), but most participants noticed at only 10\% of vertices. This, however, is a very substantial, and perhaps unrealistic, level of degradation. In addition, we do not know of an appropriate technique for scoring the quality of user-defined meshes. 
 
\textbf{Wall Material} errors were always noticed by all participants, with early noticing orders. Along with overall light color, these were perhaps the most noticeable changes, as they result in a global change across the visual field. As with other errors that were noticed by all participants, we saw no need to investigate further.
 
\textbf{Furniture Mismatch} errors were almost always noticed by all participants. This is perhaps to be expected, as the furniture objects are large and readily noticeable by participants. 
 
\textbf{Clutter Quality} errors were never noticed by any participants, even in the 10\% condition. We saw no utility in investigating this type of error further. 
 
\textbf{Light Brightness} errors were mostly not noticed by participants, even when reduced to 25\% or increased to 400\%. It appears that participants readily adapt to changes in the overall brightness of a scene, and/or consider the brightness to be independent of the room. 
 
\textbf{Light Color} errors were almost always noticed by all participants, with early noticing orders. The only error where that was not the case was when lights were interpolated 10\% toward blue. 

\subsubsection{Parameters selected for further investigation}
\label{sec:experiment1_results_furtherinvestigation}
 
\textbf{Room Length} errors were noticed by all participants at +25\% and -25\%, and by some participants at +10\% and -10\%. This suggested that there is a point of subjective equivalence somewhere between +10\% and +25\% and another between -10\% and -25\% where participants are indifferent to differences in room length. However, we did not yet know what those points are. Therefore, room length was selected for further investigation.
 
\textbf{Room Width} errors of -25\% were noticed by all participants, and all other errors were noticed by most participants. This overall greater level of noticeability compared to room length was likely due to the fact that in this room, furniture was primarily placed along the east and west walls, which were the walls that were moved to manipulate room width. We suspect that in general, either length or width will be more important depending on the primary orientation of furniture in the room. Therefore, as with room length, room width was selected for further investigation.
 
\textbf{Furniture Elevation} errors were always noticed when the furniture was moved up (off the floor), and sometimes noticed when the furniture was moved down (into the floor). We suspect this is because it is readily apparent (and non-physical) when an object is floating in the air, but it is not immediately clear when a piece of furniture goes through the floor, since it is possible that the couch doesn't have feet (for example). Also, this error is less apparent than Clutter Elevation (discussed later), because the clutter elevation errors are likely to take place at an elevation closer to eye level, while furniture elevation errors take place at floor level.
 
\textbf{Furniture Scale} errors were always noticed by participants at +25\% or -25\%, but were only noticed by some participants at +10\% or -10\%.This suggests that there is a point of subjective equivalence somewhere between +10\% and +25\% and another between -10\% and -25\% where participants will be indifferent to differences in furniture scale. However, we do not yet know what those points are. As with room length and room width, then, furniture scale was selected for further investigation.
 
\textbf{Clutter Missing} errors were almost always noticed by participants at all levels (10\% missing, 25\% missing, and 50\% missing). However, rarely, if ever, could participants identify all pieces of missing clutter and instead noticed one or a few objects (of many) that were missing. Given that there were eighty-eight pieces of clutter in the room that can individually be hidden or revealed in random orders per trial, this could conceivably be used to find a ``just noticeable difference'' in the amount of clutter that must be missing before it is generally noticeable. (This is especially true if the ``random'' orders are required to be physically accurate, which they were not in these experiments---a book could disappear from underneath another book, for example.) It has also been discussed that perhaps clutter should only be defined as small objects that are near many other small objects, so as not to remove particularly ``prominent'' objects. This is the parameter that generated the most discussion within the team. Therefore, clutter missing was selected for further investigation.
 
\textbf{Clutter Elevation} errors were almost always noticed in all the conditions. These errors are even more noticeable than Furniture Elevation errors because they generally appear closer to eye level and also because an object that interpenetrates a thin surface is more visually apparent than an object that penetrates the floor, because you can see that part of the object is below the surface. These are very noticeable errors, and therefore clutter elevation was selected for further investigation.
 
\textbf{Lights Missing} errors were mostly not noticed by participants. The room was always lit by ``ambient'' light in Unity; this seemed to be sufficient for most participants. Even lighting changes that made significant changes to the overall appearance of the room were rarely (sunlight) or never (lamplight) commented on. We suspect that this may be because the difference in lighting between the real world room and the ground truth virtual room is potentially much greater than the difference between the ground truth virtual room and the degraded virtual rooms. That is, even in our best-case virtual room, we are only capturing a fraction of the lighting effects that are present in the real room. It may be that once participants have accepted the fact of being in a computer-generated environment, they do not care much about the quality of light. Another complicating factor is that we did not compare to any ``ambient light only'' rooms; there was always at least one light that cast shadows. The data from Experiment 1, therefore, suggests that it is not worth exploring lights missing errors any further or penalizing them in the scoring system. However, it is possible that the types of lights missing errors presented in Experiment 1 are not representative of the type or scale of errors that could realistically occur in submissions. Therefore, to cover our bases, lights missing errors were selected for further investigation.
 
%
%
%

\section{Study 2}
\label{sec:experiment2}

As described in Section \ref{sec:experimentalDesign}, this user study was a psychophysical study of which the primary goal was to identify subjective equivalences between different parameters of the VE, specifically the seven parameters identified in Section \ref{sec:experiment1_results_furtherinvestigation}.
 
In order to accomplish this, we designed the study as follows. In each trial, a participant would experience three versions of the virtual replica room. First, the GT room as in Study 1, to give the participant a point of comparison. Second, an Exposure room, in which one of the seven parameters was set to one of five levels (including unchanged), and the participant was asked to verbally rate how different the Exposure room felt from the GT room, on a scale from 1 to 7. Finally, the participant experienced a Test room, in which they were able to control one of six parameters (six because the parameter from the Exposure room could not be reused), and were asked to adjust that parameter using the Oculus Touch joystick until the Test room felt ``as different'' from the GT room as the Exposure room had.
 
As an example, consider that the parameter that is varied in the Exposure room is room width, and that it was set to 50\% of the true room width. The participant might consider this room very different from the GT room, and assign a difference rating of 7. Then, in the Test room, the participant-controlled parameter is the number of clutter objects. They are asked to adjust this parameter until the Test room feels as different from the GT room as the Exposure room did; that is, until it feels like a 7. They then remove all the clutter objects from the room (setting number of clutter objects to 0), and declare a match. This ends the trial.
 
\begin{figure}[htbp]
 \centering 
 \includegraphics[width=\textwidth]{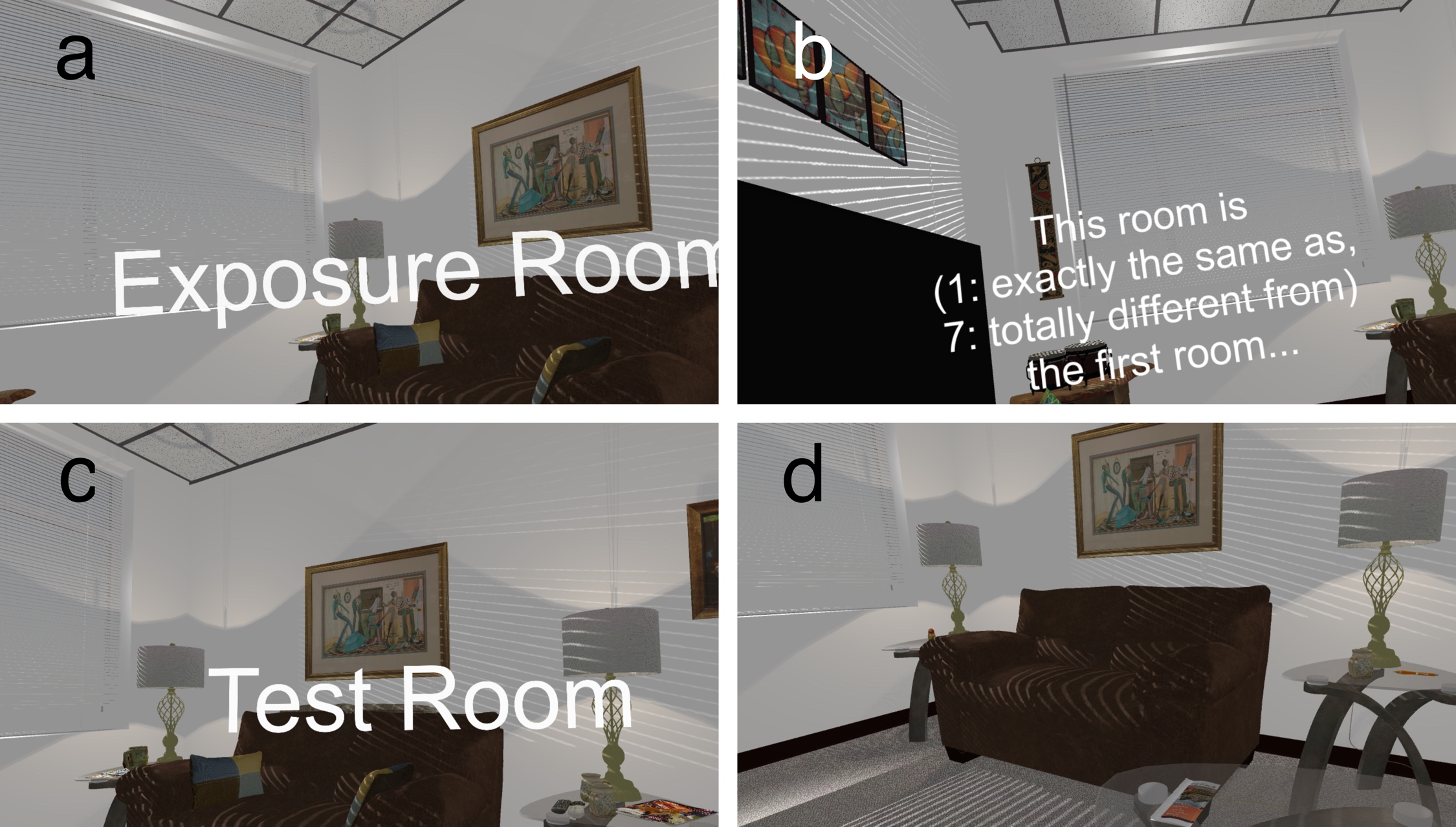}
 \caption{Screenshots from the Exposure Room (a, b) and Test Room (c, d) for one trial in Study 2.}
 \label{fig:study2_combined}
\end{figure}

Each participant underwent 210 trials (7 exposure room parameters x 5 stimulus levels of exposure room parameters x 6 test room parameters). The total duration was approximately 3-4 hours. The system automatically suggested breaks after every 20 trials (approximately every 20 minutes). Screenshots from the Exposure and Test rooms appear in Figure \ref{fig:study2_combined}.

\subsection{Participants}
\label{sec:experiment2_participants}


Eight participants (six female) were recruited from the student population of Virginia Tech's Grado Department of Industrial and Systems Engineering, in which the study was being conducted. Participants were compensated at a rate of \$5/half hour. This user study was approved by the Virginia Tech Institutional Review Board, \#17-491.
 
\subsection{Materials}
\label{sec:experiment2_materials}

The materials used in this study were the same as in Study 1, with the addition of an Oculus Touch controller. The controller was used so that the user could manipulate environment parameters by moving the joystick left and right.
 
\subsection{Measures}
\label{sec:experiment2_measures}

For each trial, two data points were collected. The first was a ``difference rating'' (on a scale from 1-7, as shown in Figure \ref{fig:study2_combined}b) of the Exposure room compared to the GT room, and the second was the ``point of equivalence'' in the Test room---that is, the level at which the participant deemed the parameter in the Test room to feel ``as different'' from the GT room as the Exposure room did.
 
\subsection{Procedures}
\label{sec:experiment2_procedures}

Upon arriving at the lab, participants read and signed an informed-consent form and completed a short demographic questionnaire. Participants were informed both verbally and in writing that they were free to withdraw from the study at any time and for any reason. After completing this process, participants donned the Oculus Rift HWD and were exposed to a familiarization environment, in which they calibrated the Rift for their IPD and familiarized themselves with moving in the Rift and the Guardian system. After completing the familiarization process, participants were began the trials, which proceeded as described in Section \ref{sec:experiment2}.
 
\subsection{Results}
\label{sec:experiment2_results}

\begin{figure}[htbp]
 \centering 
 \includegraphics[height=8.5in]{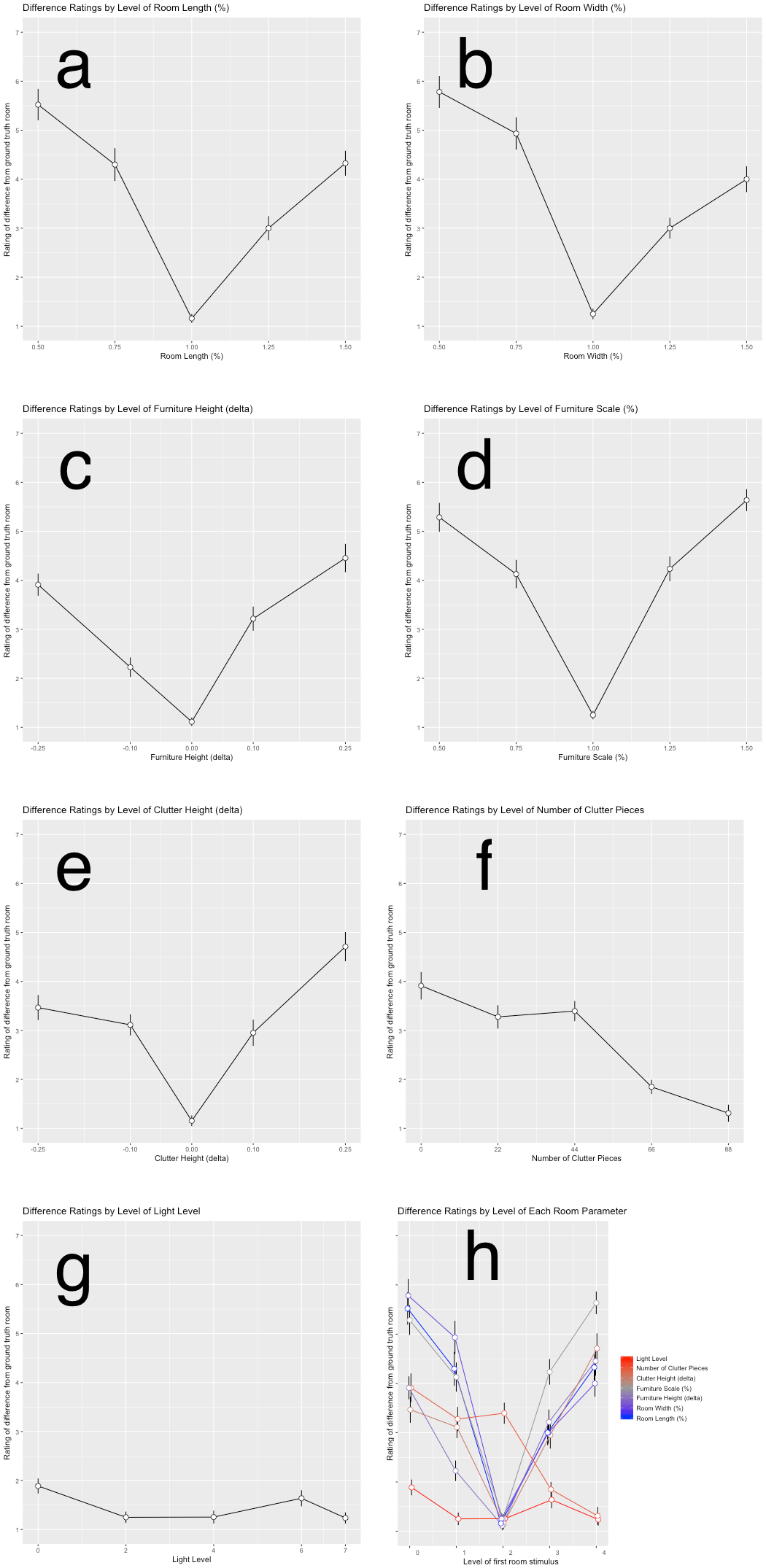}
 \caption{Graphs of difference rating by each parameter; (a) room length, (b) room width, (c) furniture height, (d) furniture scale, (e) clutter height, (f) clutter pieces, (g) lighting condition, (h) all parameters together.}
 \label{fig:study2_graphs1}
\end{figure}

\begin{figure}[htbp]
 \centering 
 \includegraphics[height=8.5in]{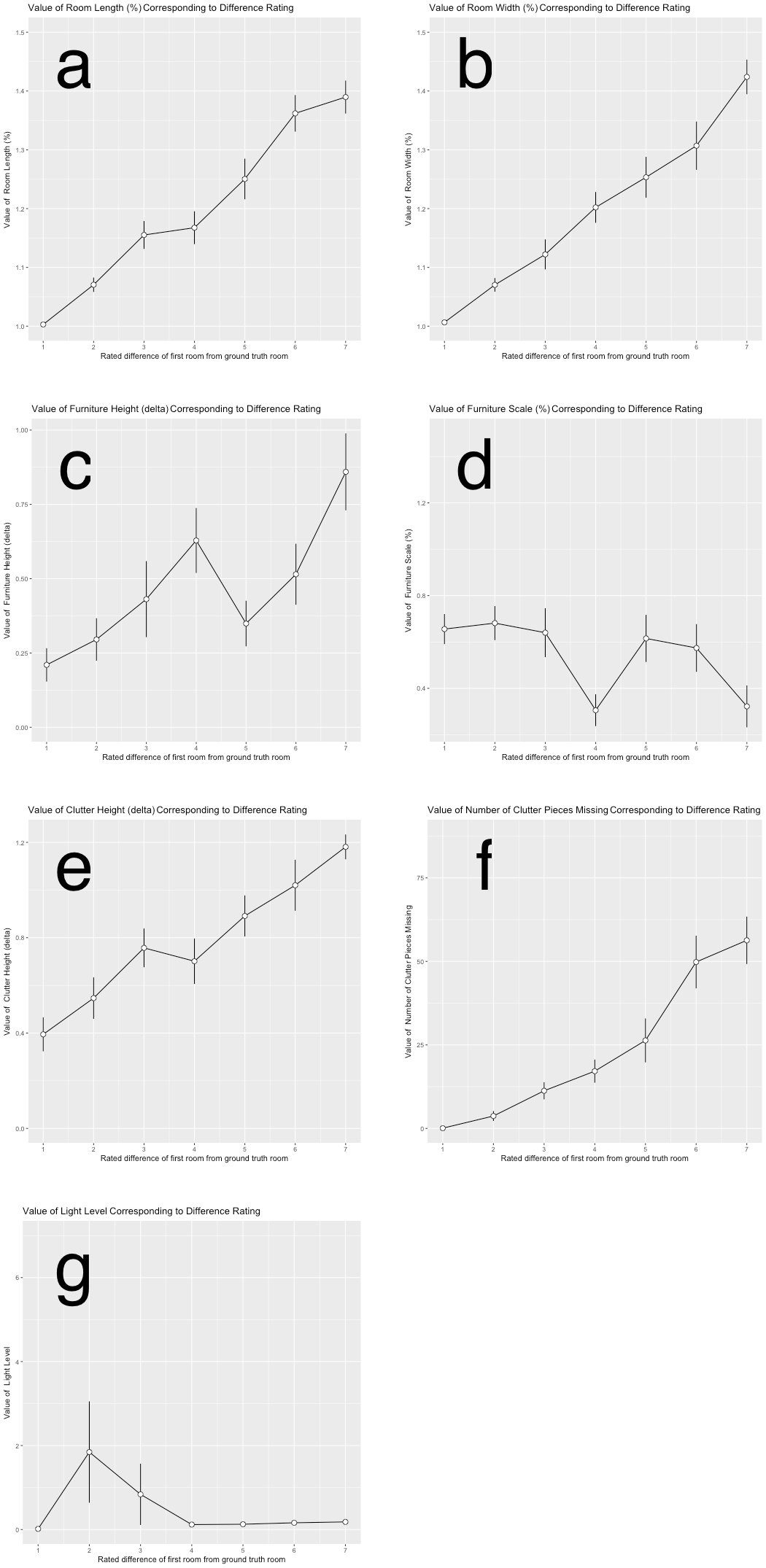}
 \caption{Graphs of the level of each parameter in the Test room corresponding to a given difference rating in the Exposure room; (a) room length, (b) room width, (c) furniture elevation, (d) furniture scale, (e) clutter elevation, (f) clutter pieces missing, (g) lighting condition.}
 \label{fig:study2_graphs2}
\end{figure}

The results from Study 2 are summarized in the graphs in Figures \ref{fig:study2_graphs1} and \ref{fig:study2_graphs2}. Figure \ref{fig:study2_graphs1} shows, for each Exposure room parameter, the resulting difference ratings at each level of that parameter. Figure \ref{fig:study2_graphs2} shows, for each Test room parameter, the Test room value that was chosen to match a given difference rating from the Exposure room.

Some of these data are discussed in greater detail in Section \ref{sec:discussion}, but to summarize, Figure \ref{fig:study2_graphs1} can be used to answer several questions regarding each parameter, such as: Did the experimental manipulation work? (Figures \ref{fig:study2_graphs1}a-\ref{fig:study2_graphs1}e indicate that for all these parameters, the rooms that were actually no different from the GT room were perceived as no different, indicating that our manipulation was effective.) Does the parameter exhibit asymmetric effects? (For example, \ref{fig:study2_graphs1}a and \ref{fig:study2_graphs1}b show that smaller rooms are perceived as ``more different'' than larger rooms, while \ref{fig:study2_graphs1}c shows that furniture height is perceived as roughly equally different whether the furniture is above or below the ground plane.) Does one parameter have a larger effect than another? (Figure \ref{fig:study2_graphs1}h shows the graphs for all parameters plotted on one set of axes; it is clear here that room width, room length, and furniture scale were the factors that resulted in the greatest perceptual difference.)

Figure \ref{fig:study2_graphs2}, on the other hand, mostly serves as a check on the results. If a parameter behaves ``as expected''---that is, such that a greater deviation from the veridical value is needed to match a greater difference rating---one would expect to see a fairly linear, monotonic increase. (This is basically what is observed in Figures \ref{fig:study2_graphs2}a, \ref{fig:study2_graphs2}b, \ref{fig:study2_graphs2}e, and \ref{fig:study2_graphs2}f, indicating that this relationship holds for room length, room width, clutter height, and clutter pieces missing.) On the other hand, furniture height, furniture scale, and lighting do not exhibit this relationship. Exploring these results is an interesting area for future work, but due to space constraints, we do not discuss Figure \ref{fig:study2_graphs2} further in this paper.

\section{Study 3}
\label{sec:experiment3}

As described in Section \ref{sec:experimentalDesign}, this user study was a ``budget-based'' study of which the primary goals were to generate a rank ordering of the studied environment parameters, as well as ``how correct'' each parameter needed to be, for the virtual replica room to feel as perceptually similar to the real original room as possible.
 
In this study, we used the same parameters from Study 2, however, the lights were broken up into three separate budget items (sun light, lamp lights, and ceiling lights), and room length and room width were combined into a single parameter, room scale. We refer to each instance of these eight parameters as a \emph{configuration}, and denote a configuration with a property vector of the form C = \{ RoomScale, FurnitureElev, FurnitureScale, ClutterPieces, ClutterElev, LampLight, SunLight, CeilingLight \}. Details regarding the costs associated with each of these parameters can be seen in Table \ref{tab:parametercosts}.

\clearpage

\begin{table*}
  \centering
  \caption{List of parameters, ranges, increments, and costs for Study 3.}
  \label{tab:parametercosts}
  \begin{tabular}{llllp{.2\textwidth}}
    \toprule
    Parameter & Initial value & Max Value & Increment & Cost in points/unit\newline (total cost) \\
    \midrule
    RoomScale & 0.5x & 1.0x & 0.01x & 1.5 (75) \\
    FurnitureElev & -25cm & 0cm & 1cm & 2 (50) \\
    FurnitureScale & 0.5x & 1.0x & 0.01 & 2 (100) \\
    ClutterPieces & 0 pieces & 72 pieces & 1 piece & 0.5 (36) \\
    ClutterElev & -25cm & 0cm & 1cm & 1 (25) \\
    LampLight & 0 (off) & 1 (on) & 1 & 10 (10) \\
    SunLight & 0 (off) & 1 (on) & 1 & 10 (10) \\
    CeilingLight & 0 (off) & 1 (on) & 1 & 10 (10) \\
  \bottomrule
\end{tabular}
\end{table*}

Each participant was first exposed to the real original room, as depicted in Figure 1a. Participants were instructed to, ``Pay attention to `how real' this room feels; you are going to experience several copies of this room in virtual reality, and we will ask you to change the virtual room until it feels as real as possible.'' After this, they were escorted to the virtual reality room, where they donned the Oculus Rift and Touch controllers, and experienced a substantially modified version of the GT room, and were given a points budget to make improvements to that room. (Upgrading every parameter to the maximum level cost 316 points, as shown in Figure 6 and broken down in Table \ref{tab:parametercosts}. In the training environment, participants were given a budget of 316 points, so as to expose the user to all the possible upgrades and what they felt like. During each recorded trial, participants worked with a restricted improvement budget of 250.) Participants then doffed the equipment, were re-exposed to the real original room, and re-donned the equipment for the next trial in virtual reality. This process was repeated for each recorded trial, of which there were seven for each participant. (Each trial started from one of the seven configurations listed in Table \ref{tab:configurationslist}; these were presented to each participant in randomized order.) Note also that each trial began with a different (randomly-chosen) parameter selected, but the parameters were always listed in the same order. So, for example, a trial could start on any of the eight parameters, but \emph{Furniture Elevation} would always appear between \emph{Room Scale} and \emph{Furniture Scale}.

\begin{table}
  \centering
  \caption{List of starting configurations for trials in Study 3.}
  \label{tab:configurationslist}
  \begin{tabular}{llp{.05\textwidth}}
    \toprule
    Starting configuration & Improvements pre-assigned & Points\newline available\\
    \midrule
    \{0,0,0,0,0,0,0,0\} & None & 250 \\
    \{1,0,0,0,0,0,0,0\} & RoomScale (75) & 175 \\
    \{0,1,0,0,0,0,0,0\} & FurnitureElev (50) & 200 \\
    \{0,0,1,0,0,0,0,0\} & FurnitureScale (100) & 150 \\
    \{0,0,0,1,0,0,0,0\} & ClutterPieces (36) & 214 \\
    \{0,0,0,0,1,0,0,0\} & ClutterElev (25) & 225 \\
    \{0,0,0,0,0,1,1,1\} & All lights on (30) & 220 \\                    
  \bottomrule
\end{tabular}
\end{table}

\begin{figure}[htbp]
 \centering 
 \includegraphics[height=\textheight]{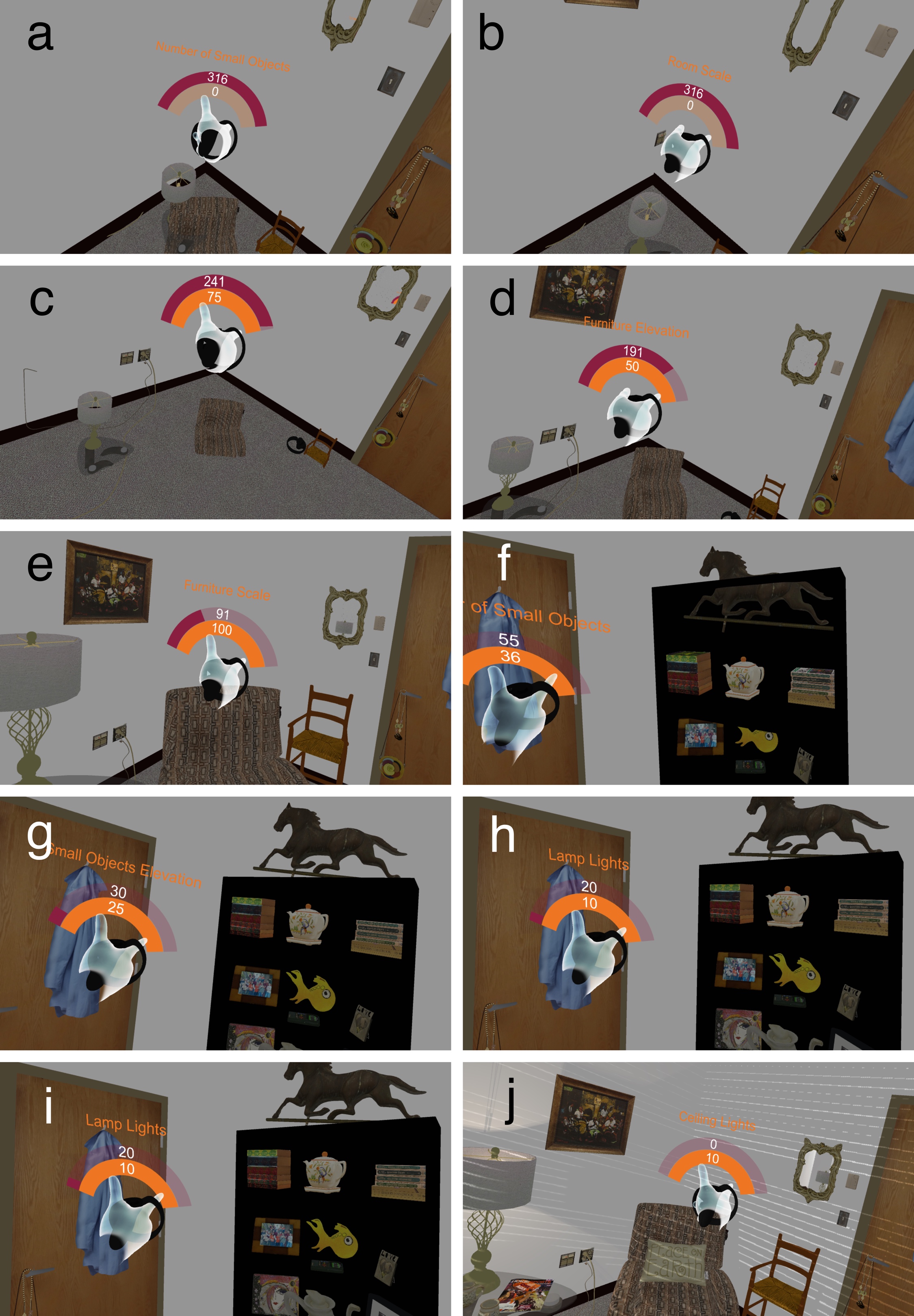}
 \caption{Screenshots showing a full sequence of all possible improvements in the Training environment of Study 3.}
 \label{fig:study3_combined}
\end{figure}

The virtual environment, along with the budget/upgrade user interface, and each of the improvements, is illustrated in Figure \ref{fig:study3_combined}.

In the virtual reality trials, each participant was instructed to, ``spend your improvement budget in order to make the room feel as `real' as possible as `quickly' as possible. That is, if one property feels most important to you, you should improve that one first, then the next most important property, and so on.'' Participants were able to explore the entire parameter space (both which parameter to adjust and how much of the points budget to spend on it), but once they confirmed their expenditure (by orally informing the experimenter), that parameter became ``locked'', and could not be revisited or further adjusted. This was done to simplify both the study procedures and the analysis. The trial ended  when the participant had spent their entire budget.

\subsection{Participants}
\label{sec:experiment3_participants}


Forty participants (nineteen female) were recruited from the student population of Virginia Tech, at which the study was being conducted. Participants were compensated at a rate of \$5/half hour. This user study was approved by the Virginia Tech Institutional Review Board, \#17-491.
 
\subsection{Materials}
\label{sec:experiment3_materials}

The materials for this study were the same as in Study 2.
 
\subsection{Measures}
\label{sec:experiment3_measures}

There were two types of dependent variable: For each trial, we recorded both (1) the sequence in which the participant chose to improve room parameters, and (2) the amount of their budget they chose to spend on each improvement. By construction, there were always an equal number of measurements of types (1) and (2), but there could be a different number of observations for each trial. Each participant underwent seven trials.
 
\subsection{Procedures}
\label{sec:experiment3_procedures}

The pre-experiment procedures here were the same as the procedures for the previous study, as described in Section \ref{sec:experiment2_procedures}. After completing the familiarization process, participants began the trials, which proceeded as described in Section \ref{sec:experiment3}. At the end of the virtual reality trials, participants completed a short post-experiment questionnaire. The entire study lasted approximately one hour.
 
\subsection{Results}
\label{sec:experiment3_results}

As in \cite{Slater:2010:SimulatingVEsInVEs}, we make the simplifying assumption that the results of the seven trials are statistically independent. This is not strictly the case; each participant carried out a series of trials, and learned about the environment parameters from trial to trial. However, the study was designed such that each trial began from a different starting configuration, with different pre-selected parameter changes and as a result, different points budgets. Because of these changes, participants would have had to reconsider their parameter choices in every trial.

In the remainder of this section, we report separately on the three types of dependent variables: the parameter sequences chosen in each trial, the amount of budget spent on each parameter, and the post-experiment questionnaire data.

\subsubsection{Transitions}
\label{sec:experiment3_results_transitions}

From the parameter sequences chosen in each trial, we constructed a transition probability matrix $P$. Over the 280 total trials, there were 1727 observed parameter changes, for an average of 6.17 parameter changes per trial. By the design of Study 3, $P$ is a 256x256 matrix. (There are 8 parameters, each of which can be in one of two states: changed (1) or unchanged (0). So there are 2\textsuperscript{8} = 256 possible configurations, and to consider the probability of a transition from any state to any other state, P needs to have 256\textsuperscript{2} cells.) Because we placed several restrictions on allowed parameter choices at any given time, P is quite a sparse matrix: there are only 273 distinctly observed state transitions (out of 65536).

Once $P$ is known, it is possible to compute the probability distribution of transitioning to any given configuration from any given configuration. If we choose as a starting configuration $C = {0,0,0,0,0,0,0,0}$ (no parameter changed), and define $s$ as a 256-vector of all zeros except the element corresponding to $C$, then $sP$ gives the probability distribution after one parameter has been changed, $sP^2$ after 2 changes, and $sP^n$ after n changes. By construction, the configuration $C = {1,1,1,1,1,1,1,1}$ is absorbing, so the eighth transition adds no information. Therefore, we report on the first seven transitions in Table \ref{tab:likelytransitions}.

\begin{table*}
  \centering
  \caption{Most likely state and next transition at every number of parameters changed (For readability, only transitions with probability $\geq$ 0.1 are shown).}
  \label{tab:likelytransitions}
  \scalebox{0.75}{
  \begin{tabular}{llllp{.2\textwidth}}
    \toprule
    Starting configuration & & Probable transitions & & Most likely action\\
    \midrule
    \{0,0,0,0,0,0,0,0\} & \{0,1,0,0,0,0,0,0\} (0.1) & \textbf{\{1,0,0,0,0,0,0,0\} (0.75)} & & Change RoomScale \\
    \{1,0,0,0,0,0,0,0\} & \textbf{\{1,0,1,0,0,0,0,0\} (0.557)} & \{1,1,0,0,0,0,0,0\} (0.357) & & Change FurnitureScale \\
    \{1,0,1,0,0,0,0,0\} & \textbf{\{1,1,1,0,0,0,0,0\} (0.859)} & & & Change FurnitureElev \\    
	\{1,1,1,0,0,0,0,0\} & \{1,1,1,0,0,0,1,0\} (0.110) & \textbf{\{1,1,1,1,0,0,0,0\} (0.764)} & & Change ClutterPieces\\    
    \{1,1,1,1,0,0,0,0\} & \textbf{\{1,1,1,1,1,0,0,0\} (0.875)} & & & Change ClutterElev \\                       
	\{1,1,1,1,1,0,0,0\} & \{1,1,1,1,1,0,0,1\} (0.195) & \{1,1,1,1,1,0,1,0\} (0.374) & \textbf{\{1,1,1,1,1,1,0,0\} (0.431)} & Change LampLight \\
	\{1,1,1,1,1,1,0,0\} & \textbf{\{1,1,1,1,1,1,0,1\} (0.5)} & \textbf{\{1,1,1,1,1,1,1,0\} (0.5)} & & Change remaining lights in either order\\   
	\bottomrule
\end{tabular}}
\end{table*}

\subsubsection{Expenditures}
\label{sec:experiment3_results_expenditures}

At each step, participants had to decide not only which parameter to change, but how much of their points budget to spend on that parameter. This data was collected in order to estimate ``how correct'' each parameter had to be. For example, participants might regard RoomScale as the most important parameter, but consider it to be ``close enough'' at 0.75, while another parameter might need to be exactly right to be acceptable. Summary statistics regarding the expenditure data appear in Table \ref{tab:expenditures}.

\begin{table*}
  \centering
  \caption{Summary statistics of expenditures.}
  \label{tab:expenditures}
  \begin{tabular}{lllllll}
    \toprule
     & & Expenditure & & & Parameter value & \\
    Parameter & Mean & Median & S. D. & Mean & Median & S. D. \\
    \midrule
	RoomScale & 61.2 & 63 & 11.4 & 0.908 & 0.92 & 0.076 \\
	FurnitureElev & 39.7 & 43 & 11.2 & -0.051 & -0.035 & 0.056 \\
	FurnitureScale & 69.3 & 70 & 18.3 & 0.847 & 0.85 & 0.091 \\
	ClutterPieces & 26.5 & 28.75 & 9.58 & 53.0 & 57.5 & 19.2 \\
	ClutterElev & 20.3 & 25 & 7.93 & -0.047 & 0 & 0.079 \\
	LampLight & 4.02 & 0 & 4.92 & 0.402 & 0 & 0.492 \\
	SunLight & 6.65 & 10 & 4.73 & 0.665 & 1 & 0.473 \\
	CeilingLight & 4.31 & 0 & 4.97 & 0.431 & 0 & 0.497 \\
	\bottomrule
\end{tabular}
\end{table*}

\subsubsection{Questionnaires}
\label{sec:experiment3_results_questionnaires}

After completing the study, all participants completed a questionnaire. We do not report data for all responses here, choosing to focus on two aspects of the questionnaire: (1) ``feeling of reality'' scores , and (2) participants' subjective rankings of the importance of the studied parameters.

Regarding their feelings of reality, participants were given two prompts: ``Picture in your mind the WORST version of the virtual room. On a scale of 0 to 100 (100 being equally as real as the real world room), how real did that room feel to you?'' and ``Picture in your mind the BEST version\ldots'' The summary statistics regarding these two scores are shown in Table \ref{tab:feelingofreality}; as one would expect, participants rated their memory of the BEST room much higher, with high statistical significance.

\begin{table}
  \centering
  \caption{``Feeling of reality'' statistics.}
  \label{tab:feelingofreality}
  \begin{tabular}{llll}
    \toprule
 	Mean(WORST) & Mean(BEST) & Difference & \\
    \midrule
	46.5 & 81.3 & 38.2 & $p < 0.001$ \\
	\bottomrule
\end{tabular}
\end{table}

Regarding their subjective rankings of parameter importance, these added credence and context for the psychophysical data presented in Sections \ref{sec:experiment3_results_transitions} and \ref{sec:experiment3_results_expenditures}. Participants were given a series of prompts, of the form ``When improving the virtual environment, which factor was most important [second-most important, third-most important, etc.] for you?'' Participants could choose any of the eight parameters, as well as ``No factor was particularly more [less] important than the others.'' These data are presented in Table \ref{tab:subjective}, similarly to Table \ref{tab:likelytransitions}. Note the similarities between these two tables. 

\begin{table*}
  \centering
  \caption{Ranking of subjectively most important parameters (For readability, only factors preferred by $\geq$ 10\% of participants are shown).}
  \label{tab:subjective}
  \scalebox{0.8}{
  \begin{tabular}{llllll}
    \toprule
 	Ranking & & & Most important factors & & \\
    \midrule
	1\textsuperscript{st} & \textbf{RoomScale (67.5\%)} & FurnitureScale (15\%) & & & \\
	2\textsuperscript{nd} & RoomScale (17.5\%) & FurnitureElev (10\%) & \textbf{FurnitureScale (70\%)} & & \\
	3\textsuperscript{rd} & RoomScale (10\%) & \textbf{FurnitureElev (40\%)} & FurnitureScale (15\%) & ClutterPieces (20\%) & \\
	4\textsuperscript{th} & FurnitureElev (15\%) & \textbf{ClutterPieces (40\%)} & ClutterElev (30\%) & NONE (10\%) & \\
	5\textsuperscript{th} & FurnitureElev (10\%) & ClutterPieces (25\%) & \textbf{ClutterElev (37.5\%)} & SunLight (15\%)  & NONE (10\%) \\	
	6\textsuperscript{th} & FurnitureElev (10\%) & LampLight (15\%) & \textbf{SunLight (40\%)} & CeilingLight (10\%)  & NONE (20\%) \\	
	7\textsuperscript{th} & \textbf{LampLight (30\%)} & SunLight (17.5\%) & CeilingLight (17.5\%) & \textbf{NONE (30\%)} &  \\
	8\textsuperscript{th} & LampLight (25\%) & SunLight (12.5\%) & \textbf{CeilingLight (42.5\%)} & NONE (20\%) &  \\				
	\bottomrule
\end{tabular}}
\end{table*}

\section{Discussion}
\label{sec:discussion}

Throughout this section, we discuss the implications of the results from the three studies organized as claims about the data followed by the supporting evidence for those claims.

\subsection{Manipulating room parameters in this way does in fact change users' ``feeling of reality''}
\label{sec:discussion_1}

As mentioned in Section \ref{sec:experiment3_results_questionnaires}, participants asked to remember the best version of the virtual replica room rated it more real than the worst version of the replica room. This is not necessarily surprising, but it does validate our experimental manipulations. Changing room parameters in this way affects the coherence of the virtual environment, and thereby seems to affect users' feelings of Plausibility Illusion (Psi) in that environment, as anticipated.

\subsection{Room Scale is the most impactful of the studied parameters}
\label{sec:discussion_2}

In both Study 2 (in which room length and room width at 0.5x represented 2 of the 3 highest observed difference ratings over the whole parameter set) and Study 3 (in which room scale was subjectively the most important parameter to a substantial majority of participants, was the first parameter upgraded in a substantial majority of trials, and had the second highest gross mean expenditure), room scale was observed to be the most impactful parameter.

This is perhaps not surprising, as the scale of the room overall provides the context by which to evaluate the scale of individual objects. (Correctly-sized furniture in an implausibly small room would likely appear even more incoherent than implausibly small furniture in the same room.) From Study 2, we can see that participants rated smaller rooms as more incoherent than larger rooms for the same GT room (Figures \ref{fig:study2_graphs2}a and \ref{fig:study2_graphs2}b). In Study 3, the median accepted room scale was 0.92x, indicating that participants were willing to accept a room approximately 10\% smaller than the real room as “feeling real.”

\subsection{Furniture Scale is the second most impactful of the studied parameters}
\label{sec:discussion_3}

In both Study 2 (in which furniture scale of 1.5x represented the second highest overall difference rating) and Experiment 3 (in which furniture scale was subjectively the second important parameter to a substantial majority of participants, was the second parameter upgraded in a substantial majority of trials, and had the highest gross mean expenditure), furniture scale was observed to be the most impactful parameter.

Again, this is not particularly surprising. Furniture objects are the biggest objects in the room (save for the room itself), so might be expected to make the biggest impact on the coherence of the room. In Study 3, the median accepted furniture scale was 0.85x, indicating that participants were willing to accept furniture 15\% smaller than GT as ``feeling real.''	 

Note that this is in a room that was accepted as 10\% smaller than GT on average, so 15\% overstates the difference. If one considers room scale and furniture scale to be a single percept of ``relative scale'', the relative scale difference between the two is only 5\%. Study 2 provides some evidence that room scale and furniture scale are linked in this way, as while the room scale differences were perceived as worse when the room was smaller than GT, the furniture scale differences were perceived as slightly worse when the room was bigger than GT (Figure \ref{fig:study2_graphs2}d). In both cases, the furniture was ``too big'' for the room. (And in both cases, this was perceived as less realistic than the furniture being ``too small'' for the room.)

\subsection{Participants have low tolerance for non-physical behavior}
\label{sec:discussion_4}

Furniture elevation and clutter elevation were different from the other parameters in this study, in that they explicitly created non-physical environments (i.e., objects sunken into the floor, objects floating in air, objects interpenetrating one another). These parameters were perceived as less important than both scale parameters, ranking 3rd and 5th, respectively, in both subjective rankings and transition probability in Study 3. That said, participants had very little tolerance for error in either of them, as can be seen in Table \ref{tab:expenditures}. In particular, the median expenditure on clutter elevation was 25 out of 25 possible points, indicating that in a majority of trials, participants chose to spend their budget to remove all error from clutter elevation. (Participants generally chose to leave some error in furniture elevation, but we believe this can be at least partially explained by the fact that, as mentioned in Section \ref{sec:experiment1_results_furtherinvestigation}, furniture elevation errors present at ground level, while clutter elevation errors present closer to eye level.)

\subsection{Lights were considered to be not impactful, or even negatively impactful}
\label{sec:discussion_5}

Across all metrics, across Studies 2 and 3, the lighting parameters were last and least upgraded, and subjectively considered least important. We suggest three possible explanations for this. First is that in the real world, people adapt to wildly differing and rapidly changing lighting conditions with little difficulty. Despite occasional misperceptions, such as the ``dress color illusion,'' we are generally able to perceive colors as stable even as lighting changes. It is possible that we have learned and/or evolved in such a way as to be able to separate (and ignore) the effects of lighting in favor of a stable world model. Second is that despite great advances, real time computer-generated lighting is still unable to capture many of the effects of lighting in the real world. It may be that even the ``best'' lighting conditions in these user studies were perceived as low-quality by participants, so they did not bother to spend time or points upgrading the lighting. Finally, the lighting conditions in the studies did not exactly match the lighting conditions in the real room. (This is discussed further in Section \ref{sec:futurework}.) If participants were trying to exactly match the conditions they experienced in the real original room, rather than trying to match a ``feeling of reality'' in a broader sense, it makes sense that they would not have chosen to turn lights on if lighting conditions remained ``non-matching.''

\subsection{Absolute scale is more salient---and therefore more important---in VR than in other types of media}
\label{sec:discussion_6}
Section \ref{sec:discussion_3} discussed the importance of the \emph{relative} scale of objects in a virtual replica room, as users seem to be keenly aware of the fact that an object is too big or too small compared to its context. However, \emph{absolute} scale is also critically important in virtual reality experiences, in a way that is not the case with other media. In a single image or even a movie (particularly one that is computer generated), it is impossible to judge the absolute scale of the scene, since the camera is just a point, can have any height, and can move at any speed. (Consider the historically widespread use of miniatures in filmmaking.) In VR, however, the user themself provides context for absolute scale: whether or not the user is ``embodied'' in the sense that there is a tracked representation of the user visible in the virtual experience, they are still embodied in the sense that the rendered imagery is deeply and inherently connected to the human body. (Consider configuration parameters such as eye height and IPD, as well as dynamic characteristics such as the fact that camera movement is now limited by human movement constraints.) In short, VR enables a user to ``feel'' the scale in a way that even 3D content does not.

\subsection{Why do scale changes seem to have the biggest impact on Plausibility Illusion?}
\label{sec:discussion_7}

As discussed in \ref{sec:discussion_2} and \ref{sec:discussion_3}, the most impactful parameters of those studied were room scale and furniture scale (or, perhaps, relative scale).  Section \ref{sec:discussion_6} made the case that absolute scale is also especially important for VR scenes.Less impactful were furniture elevation, clutter pieces, and clutter elevation, despite the fact that these parameters could cause obviously \emph{non-physical room states} (e.g., objects floating in air).

We propose two possible explanations for this. One is that changing the scale parameters has the largest possible sensory impact on the user (the largest objects in the room are the room shell components---walls, floor, ceiling---followed by the furniture objects). The other is that the scale parameters are subjectively more important because they are more important from an affordance perspective. That is, the scale parameters inform action possibilities---Is this chair the right size to sit in? Is that door big enough to walk through?---in a way that the other parameters do not. (We call these the \emph{sensory impact} and \emph{affordance} hypotheses, respectively.)

\section{Limitations and Future Work}
\label{sec:futurework}

Despite the progress represented by the results discussed in the previous section, there are several significant limitations that must be addressed in future work in this area. First, there are simply many factors that one could imagine would affect the design of and user experience in a virtual space that were not studied in this work. Due to the logistical constraints associated with running these user studies, we always planned to include no more than eight parameters in the studies that became Studies 2 and 3. (We ultimately included only seven parameters in these studies, based on our observations in Study 1.) 

Particularly, this work to date has only concerned itself with the perceived realism of static scenes. We explicitly did not attempt to measure the effects of object or character behavioral realism, which are certainly factors that impact a user's feeling of reality. In fact, existing research regarding coherence and Plausibility Illusion has focused almost entirely on behavioral coherence \cite{Bergstrom:2017:StringQuartet} \cite{Skarbez:2017:PsychophysicalExperiment} \cite{Slater:2010:SimulatingVEsInVEs}. Integrating that work with our research is a very interesting avenue of future work, as the community works toward a more complete model of coherence.

Perhaps a more fundamental limitation is that many of these parameters might be inherently inseparable. (Consider the discussion of room length/room width/furniture scale possibly all representing a ``relative scale'' construct in Section \ref{sec:discussion_3}.) A piece of furniture is neither realistic nor unrealistic in a vacuum; it receives spatial context, visual context, and use context from the room shell and other objects.

Another limitation of these studies is that, for feasibility reasons, in Study 3 we only considered ``one-sided'' errors. That is, all parameters were restricted to be strictly less than or equal to the veridical value. Even in this paper, it is clear that this restriction is not entirely valid; see Figure \ref{fig:study2_graphs1} for evidence that some parameters influence the sense of reality asymmetrically, such that less-than errors are more perceptually disturbing than greater-than errors, or vice versa. Allowing for both less-than and greater-than errors will significantly complicate the analysis, but this may be a necessary sacrifice to achieve improved validity.

We mentioned in Section \ref{sec:discussion_5} that the lighting conditions in the real room did not exactly match the lighting conditions available in the virtual room. When participants experienced the real room, the room was lit by the ceiling lights, as well as whatever natural light was coming through the blinds. This differs in at least three (potentially) important ways from the lighting conditions in the virtual room. First, participants never saw the real room lit with the lamps turned on. Second, the virtual ceiling lights did not actually light the room; turning on the ceiling lights in any of the studies only caused them to change color. This is because we were not able to take advantage of area lights in Unity due to our need to manipulate the models in real time. And finally, participants experienced the real room on different days, at different times of day. The virtual sun light did not change; it always appeared as bright mid-afternoon sunlight. Any or all of these may be worth revisiting as capabilities change.

Despite these limitations regarding the evaluation of lighting in our studies, it would be interesting to consider why our participants did not consider lighting to be important for their feeling of reality. In Section \ref{sec:discussion_5}, we put forward three potential explanations. Evaluating these explanations is a rich area for further work.

Similarly, in Section \ref{sec:discussion_6}, we proposed sensory impact and affordance hypotheses for why scale seems to be the most important percept for users to have a strong feeling of reality. Evaluating these hypotheses is also an avenue for future work.

Finally, we believe that it would be valuable to reach out to communities that have a richer knowledge base regarding users' perception of space, particularly psychology and architecture. It is likely that they have models that could inform future experimentation; it is also likely that using virtual replica environments as testbeds could help those fields advance as well. 

\section{Conclusion}
\label{sec:conclusion}

In this paper, we have presented the designs of and results from three user studies investigating how users perceive virtual replica environments, motivated by the emerging use case of social VR. These results represent a first attempt to measure which characteristics of a virtual space are most perceptually important to users, and include the fact that the scale of room components and large objects such as furniture are more important to user experience than other factors, such as lighting, which does not play a large role in users' feeling of reality in a virtual environment.

This work is far from finished, as evidenced by the discussion in the previous section. However, we are hopeful that the results included here can serve as the basis for and stimulate additional research into the specific issues faced when creating virtual replicas of real spaces.

\section*{Acknowledgements}
The authors wish to thank Deborah Asabere, Shabi Mustafa, and Ahmed Salih for their considerable help in running these experiments. We would also like to thank our participants, without whom this work would not have been possible. This work was supported by a grant from Facebook.

\bibliographystyle{unsrt}
\bibliography{}

\end{document}